# DPE-NET: DUAL-PARALLEL ENCODER BASED NETWORK FOR SEMANTIC SEGMENTATION OF POLYPS

Malik Abdul Manan
Faculty of Information Technology
*Beijing University of Technology*
Beijing, China
manansandila@yahoo.com

Feng Jinchao
Faculty of Information Technology
*Beijing University of Technology*
Beijing, China
fengjc@bjut.edu.cn

Shahzad Ahmed
Faculty of Information Technology
*Beijing University of Technology*
Beijing, China
shahzadahmad@cuisahiwal.edu.pk

Abdul Raheem
Faculty of Information Technology
*Beijing University of Technology*
Beijing, China
raheemabdul@emails.bjut.edu.cn

*Abstract*— **In medical imaging, efficient segmentation of colon polyps plays a pivotal role in minimally invasive solutions for colorectal cancer. This study introduces a novel approach employing two parallel encoder branches within a network for polyp segmentation. One branch of the encoder incorporates the dual convolution blocks that have the capability to maintain feature information over increased depths, and the other block embraces the single convolution block with the addition of the previous layer's feature, offering diversity in feature extraction within the encoder, combining them before transpose layers with a depth-wise concatenation operation. Our model demonstrated superior performance, surpassing several established deep-learning architectures on the Kvasir and CVC-ClinicDB datasets, achieved a Dice score of 0.919, a mIoU of 0.866 for the Kvasir dataset, and a Dice score of 0.931 and a mIoU of 0.891 for the CVC-ClinicDB. The visual and quantitative results highlight the efficacy of our model, potentially setting a new model in medical image segmentation.**

## I. INTRODUCTION

Polyps are abnormal growths that arise in various body parts but are most commonly found in the colon. Early and accurate detection of polyps, especially in the colorectal region, is crucial, as some can develop into malignant tumors [1]. Traditional manual screening procedures, though effective, are time-consuming and subject to human error. Thus, there's a pressing need for more automated and reliable detection methods, particularly in medical imaging.

CNNs have revolutionized the domain of medical image analysis [2-23], the methods designed explicitly for polyp segmentation, such as SFA [24], UNet++ [25], MSNet [26], PraNet [27], Polyp-PVT [28], MSRF-Net [28], MEGANet [29], DBMF [30], cascade attention decoding [31], CFANet [32], CoInNet [33] and SANet [34] rely on the integration of encoded feature maps from different scales and semantically rich decoded feature maps. However, as medical imaging datasets grow in complexity and size, there's a rising demand for more computationally efficient architectures that can maintain or enhance the segmentation accuracy. The contemporary research landscape has seen a proliferation of novel CNN architectures aiming to achieve better accuracy and computational efficiency. One of the promising advances is introducing blocks within the network and modular components [35] that are used to optimize specific functionalities of the CNN. The former aids in retaining the original feature information as the data propagates through layers, and the latter introduces necessary variations and complexities to capture diverse patterns in the data. The key contributions of our research include:

- We introduce a Dual parallel encoder network (DPE-Net) that facilitates a balance between depth and feature preservation. DPE-Net effectively addresses the common challenge of feature degradation in deep learning (DL) by integrating dual and single convolution blocks in parallel branches. As the network depth increases, it remains capable of capturing essential features without significant loss of information.
- Parallel encoder branches are seamlessly merged, leveraging parallelism to aggregate and enhance the representative power of the extracted features. By running these blocks in parallel and combining them before transpose layers with a depth-wise concatenation operation, we aim to harness the strengths of both blocks.

This architecture promises to enhance polyps' segmentation accuracy and expedite the computational process, ensuring that medical professionals can make quicker and more informed decisions.

## II. PROPOSED METHODOLOGY

In the rapidly evolving domain of medical image analysis, segmentation accuracy and computational efficiency are paramount pillars. The research introduces a novel DPE-Net uniquely designed for efficient polyp segmentation to merge the intrinsic strengths of dual parallel encoder blocks used for feature extraction [36]. This architecture enhances segmentation accuracy and optimizes computational efficiency with only 3.4 million parameters, bridging the gap between intricate medical image analysis and real-time clinical application. In this section, we delineate our proposed architecture's key components and operational mechanisms, shedding light on the rationale behind the design choice and its subsequent implications for polyp segmentation.

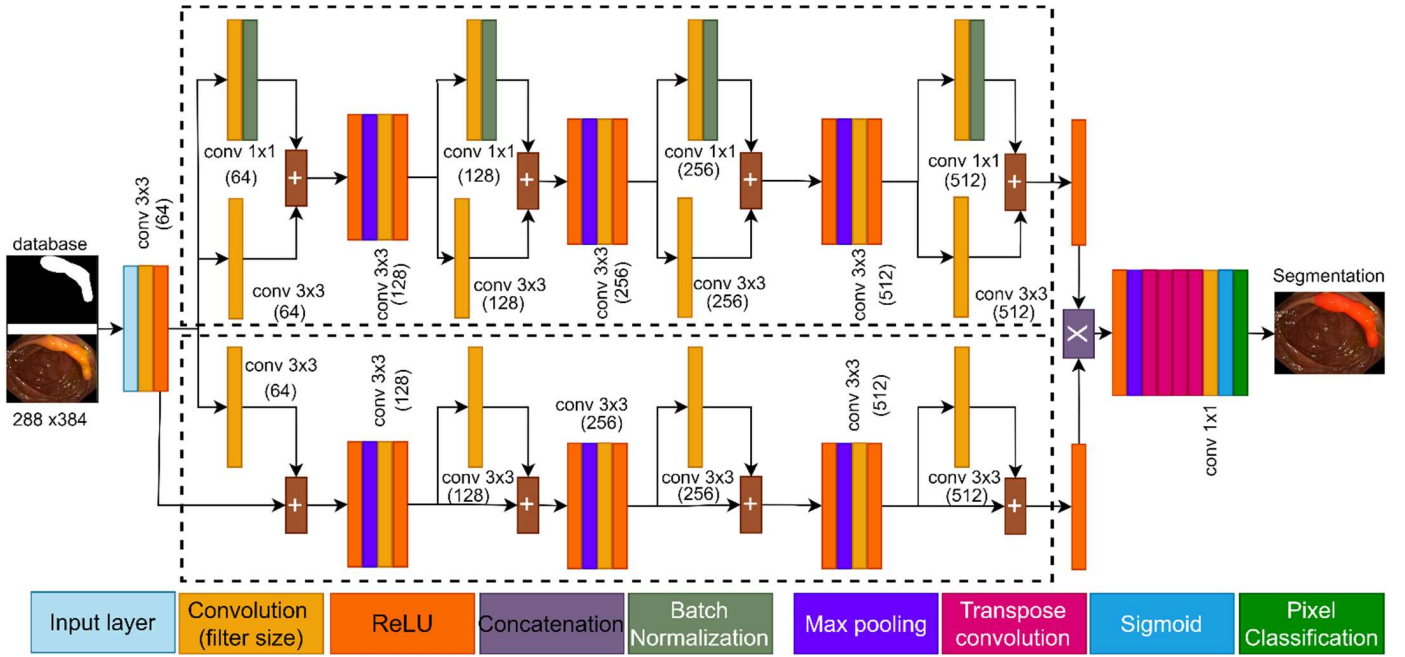

Fig. 1. The detailed architecture diagram based on a dual encoder-based Network for polyp's segmentation.

As depicted in Fig 1, the proposed model utilizes two distinct parallel encoder-based CNN branches: one based on the dual convolution block and the other on the single convolution block repeated throughout the encoder. These branches run parallel, ultimately converging before the transpose layer using a depth-wise concatenation layer. This design harnesses the strengths of both dual and single parallel convolution blocks, optimizing segmentation performance for polyps. The model accepts images of size 288 x 384, the same as the original database (Clinic DB [37]) of endoscopic images. The database comprises colorectal endoscopy images, highlighting polyps in various conditions and scenarios.

### A. Dual Convolution-Based Encoder Block

The dual convolution block-based encoder uniquely utilizes two convolution layers to introduce specific and deliberate transformations to the input data. The primary motivation behind employing these blocks is their ability to capture intricate details in the images, which in turn contributes to achieving higher precision in the segmentation of polyps. Delving deeper into the structure of this branch, it consists of several layers, each serving a distinct purpose. Initially, the convolutional layers, equipped with 1x1 and 3x3 filters, play a vital role in extracting spatial features and patterns from the input images. Following the convolutional operations, the ReLU activation function is employed. The dual convolution parallel mapping is used when the number of channels in the input feature $I_1$ does not match the channels after the previous residual function FR. We perform convolutions (1×1) for the element-wise addition.

$$I_{k+1}=M+S \quad (1)$$

Where M is the mapping of the input features map($FI_1$), $I_{k+1}$ is the output features, k represents the number of previous residual blocks, and conv (1x1) with Batch normalization produces S features, as shown in Fig 2.

### B. Single Convolution-Based Encoder Block

The single convolution block in the encoder branch of the DPE-Net is distinctively designed around the concept of Identity blocks. These blocks have the inherent characteristic of preserving the authenticity and originality of the input data, even as it undergoes the requisite transformations. This balance ensures that the inherent properties of the data remain largely untouched while still allowing the network to harness the necessary features for accurate segmentation.

$$I_{k+1}=FI_1+I_1 \quad (2)$$

Here, $I_1$ is the input features as well as element-wise addition features, $FI_1$ features after the convolution block and $I_{k+1}$ is the output features after the residual operation. A closer examination of the network's layers reveals convolutional layers exclusively employing 3x3 filters. These filters adeptly capture spatial features from the images, forming the foundation of the feature extraction process. After the convolution operations, the ReLU activation function is invoked, infusing the model with the required non-linearity, enabling it to interpret and learn more intricate patterns in the dataset.

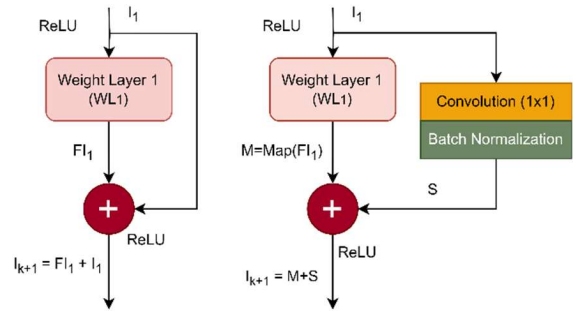

Fig. 2. The single and dual convolutions block repeatedly used in the DPE network.

Batch normalization is smoothly employed within the branch to enhance the training process. This technique optimizes the layer activations, ensuring a more stable and swift training phase. Complementing these components are the max-pooling layers, adept at down-sampling the feature maps. These layers accentuate the most significant features by reducing their spatial dimensions, refining the extraction process, and ensuring a more streamlined and precise segmentation outcome.

### C. Concatenation Approach for Polyp Segmentation

After each branch's feature extraction and necessary transformations, their respective outputs are combined using a concatenation Layer. This layer ensures a depth-wise concatenation of fine-tuned weight, effectively merging features from both branches. After this depth-wise concatenation, a Transpose Convolution layer up-samples the merged feature map, setting the stage for the ultimate segmentation phase. In the final step, a sigmoid layer is employed to classify each pixel, distinguishing the polyps from the background. This approach harmonizes the capabilities of both dual convolution block features and single convolution block in a parallel design, aiming to refine polyp segmentation in colorectal endoscopy images.

### III. EXPERIMENTAL RESULTS AND DISCUSSION

In this section, we outline the setup and findings of our study, aimed at evaluating the performance of our proposed model. The experiments were conducted on a system equipped with 16GB RAM and an NVIDIA GeForce RTX 3060 GPU, utilizing MATLAB for implementation. The network optimization was achieved through Stochastic Gradient Descent with Momentum (SGDM), with a mini-batch size of 8 and a momentum setting of 0.9. For the computational efficiency in our polyp analysis framework, we standardized the image resolution to 384x288 pixels. This uniformity helps streamline the processing pipeline. The training regimen was structured over 40 epochs, initiating with a learning rate 1e-4. The DPE-Net enhances segmentation accuracy and optimizes computational efficiency with only 3.4 million parameters, by employing techniques such as optimized batch processing, and possibly leveraging mixed-precision training with cross entropy loss, we curtail the training duration to 6 hours. These measures target a reduced computational complexity and strive to minimize training time without compromising the model's performance.

### A. Datasets and Performance Measures

We conducted experiments using two datasets, CVC-ClinicDB [37] and Kvasir [38], to assess the segmentation performance of our proposed method. The Kvasir database contains 1,000 images, and the CVC-ClinicDB with a collection of 612. Mean intersection over union (mIoU) and the mean dice similarity coefficient (mDice) measures are often employed in contests to gauge the relative performance of various models. Specifically, mIoU frequently used in medical image segmentation tasks.

$$\text{Dice} = \frac{2TP}{2TP + FN + FP} \quad (3)$$

$$\text{IoU or Jaccard} = \frac{TP}{(FP + TP + FN)} \quad (4)$$

### B. Visual Results and Comparisons with SOTA

In our study, the datasets were split using an 80-10 ratio for effective model training and testing, and the remaining 10% was used for validation purposes. Our proposed model's performance is compared against several established State-of-the-Art (SOTA) DL models. These include UNet [39], SegNet [40] with VGG16, Deeplabv3+ [41] using with ResNet 50 backbones, PraNet [27], Polyp-PVT [28], and MEGANet [29]. Fig. 3. presents a side-by-side visual comparison, showcasing the segmentation results from each model against the ground truth. The effectiveness of our proposed method is highlighted, especially when overlaid on the ground truth (GT), elucidating the precise capture of intricate details and contours.

Fig. 4. visually compares the segmentation results on the CVC-ClinicDB dataset samples using several renowned DL models. Each row illustrates a different sample from the dataset, and each column showcases the results produced by a specific model. Starting with the most popular segmentation model the UNet, SegNet and Deeplabv3+ models with 50 backbones. The PraNet [27], Polyp-PVT [28], and MEGANet [29] models

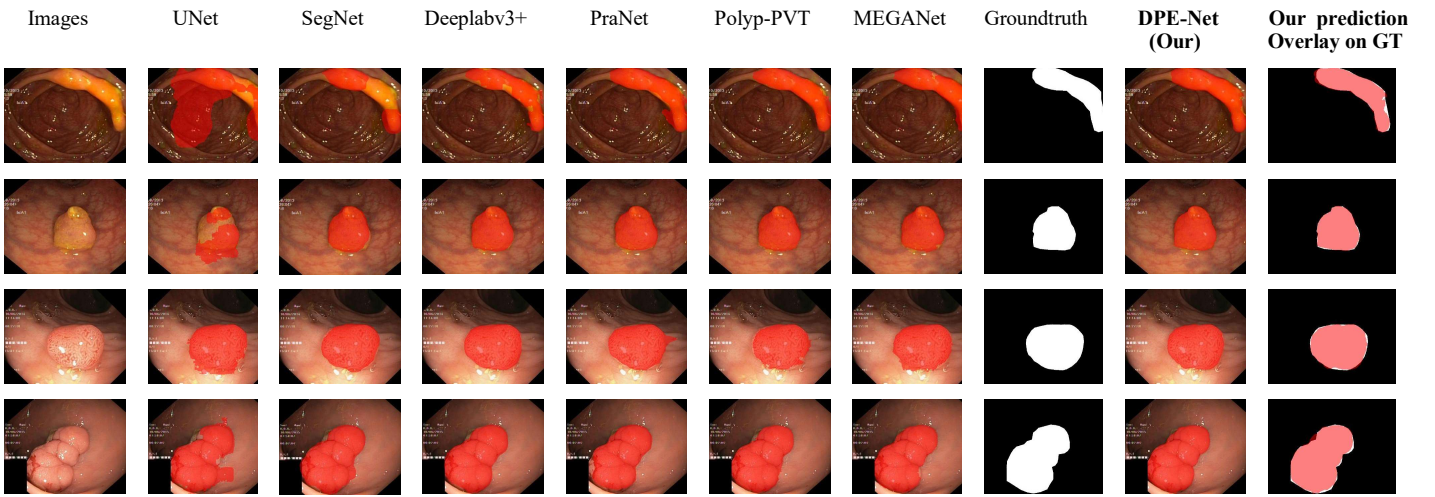

| Images | UNet | SegNet | Deeplabv3+ | PraNet | Polyp-PVT | MEGANet | Groundtruth | **DPE-Net (Our)** | **Our prediction Overlay on GT** |

Fig. 3. The comparison of visual results with deep learning models and our proposed method on Kvasir dataset.

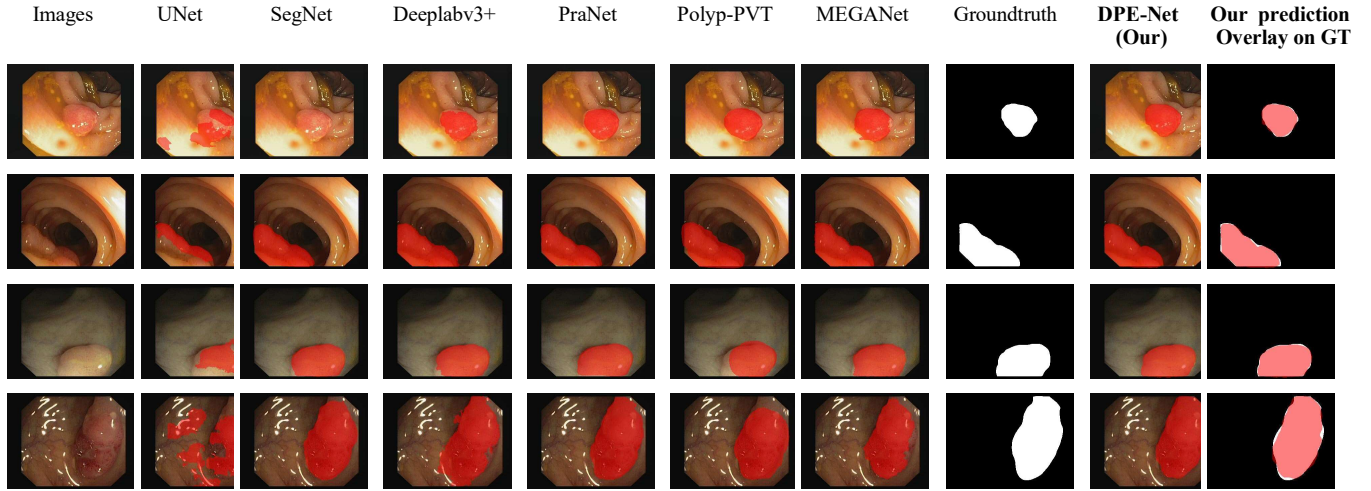

Fig. 4. The comparison of segmentation results on CVC-clinic DB dataset, including our method and various deep learning models.

accurately capture the target region's shape and contours. However, our method stands out, with the segmented region closely matching the ground truth, showing both the accuracy in boundary delineation and the completeness of the segmented region.

*C. Ablation study*

The ablation study in Table I evaluates the performance of various configurations of our model on the Kvasir and CVC-ClinicDB datasets, underscoring the contributions of individual network components to overall model efficacy. Network 1, incorporating a dual convolution block, has moderate mDice and mIoU metrics scores. A shift to a single convolution block in Network 2 reduced performance metrics. However, a significant performance boost was evident in Network 3, which integrated a dual convolution block with an optimized learning rate of $10^{-3}$, marking a significant advancement. These enhancements conclude that our proposed model, DPE-Net, outperformed the others regarding mDice and mIoU scores on both databases, underscoring DPE-Net's enhanced capability for analyzing colonoscopy images effectively.

TABLE I. THE ABLATION STUDY OF THE PROPOSED MODEL ON KVASIR AND CVC-CLINICDB DATABASES WITH TRAINING ACCURACY AND DICE.

| Models (with blocks) | Databases | | | |
|---|---|---|---|---|
| | Kvasir | | CVC-ClinicDB | |
| | mDice | Accuracy | mDice | Accuracy |
| Network 1 (Dual convolution block only) | 0.640 | 0.913 | 0.552 | 0.964 |
| Network 2 (Single convolution block only) | 0.632 | 0.91 | 0.475 | 0.948 |
| Network 3 (Both parallel blocks with LR $10^{-3}$) | 0.901 | 0.921 | 0.898 | 0.918 |
| **Our Proposed DPE-Net** | **0.919** | **0.971** | **0.931** | **0.971** |

*D. Discussion*

An examination of the performance metrics across various DL models on the Kvasir and CVC-Clinic DB databases reveals significant variations in effectiveness, particularly when evaluated through the mDice and mean mIoU. Table II shows that different deep learning models exhibit varying performance. On the Kvasir database, while the UNet and SegNet models exhibit decent performances, the Deeplabv3+ with a ResNet 50 backbone stands out with higher mDice and mIoU values from UNet and segNet models. However, the PraNet, Polyp-PVT and MEGANet model surpasses other models, coming very close to our proposed model. Our proposed model still leads the pack, showcasing the highest mDice and mIoU values. For the CVC-ClinicDB database, MEGANet yields impressive results.

Contrary to its performance on Kvasir, MEGANet seems to lag behind our DPE model. The versions of Deeplabv3+ and SegNet demonstrate competitive results compared to the Unet model. The performance of PraNet, MEGANet and Polyp-PVT models are consistent, with slight variations in performance measures.

TABLE II. THE QUANTITATIVE RESULTS OF DEEP LEARNING MODELS ON KVASIR AND CVC-CLINICDB DATABASES.

| Database | Kvasir | | CVC-ClinicDB | |
|---|---|---|---|---|
| Models | mDice | mIoU | mDice | mIoU |
| UNet [39] | 0.822 | 0.783 | 0.896 | 0.883 |
| SegNet (VGG 16) [40] | 0.844 | 0.838 | 0.654 | 0.767 |
| Deeplabv3+(ResNet 50) [41] | 0.899 | 0.915 | 0.722 | 0.741 |
| PraNet [27] | 0.898 | 0.840 | 0.899 | 0.849 |
| Polyp-PVT [28] | 0.917 | 0.864 | 0.937 | 0.863 |
| MEGANet [29] | 0.911 | 0.859 | 0.930 | 0.885 |
| **Our Proposed DPE-Net** | **0.919** | **0.866** | **0.931** | **0.891** |

## IV. CONCLUSION

This study presents an innovative approach to polyp segmentation in endoscopy images by synergizing parallel dual and single convolution block-based encoders for feature extractions. The DPE-Net capability maintains feature information over increased depths and embraces the single and dual convolution blocks, offering feature extraction diversity and employing varying block mechanisms. Our method proved efficient for capturing intricate details and achieving higher precision in publicly available Kvasir and CVC-ClinicDB databases. DPE-Net sets a new benchmark for future research and development in deep learning and computer vision by addressing key challenges such as diversity in shapes, texture and size of polyps.

ACKNOWLEDGMENTS

The authors would like to thank the sponsored organizations. This work has been sponsored by the National Natural Science Foundation of China under Grant No. 82171992 and the Beijing Laboratory of Advanced Information Networks.

REFERENCES

1. Garg, R., M. Gaur, B.G. Prajapati, G. Agrawal, S. Bhattacharya, and G.M. Elossaily, *Polymeric nanoparticles approach and identification and characterization of novel biomarkers for colon cancer.* Results in Chemistry, 2023: p. 101167.
2. Iqbal, S., A. N. Qureshi, J. Li, and T. Mahmood, *On the analyses of medical images using traditional machine learning techniques and convolutional neural networks.* Archives of Computational Methods in Engineering, 2023. **30**(5): p. 3173-3233.
3. Iqbal, S., T.M. Khan, S.S. Naqvi, A. Naveed, and E. Meijering, *TBConvL-Net: A hybrid deep learning architecture for robust medical image segmentation.* Pattern Recognition, 2025. **158**: p. 111028.
4. Matloob Abbasi, M., S. Iqbal, K. Aurangzeb, M. Alhussein, and T.M. Khan, *LMBiS-Net: A lightweight bidirectional skip connection based multipath CNN for retinal blood vessel segmentation.* Scientific Reports, 2024. **14**(1): p. 15219.
5. Farooq, H., Z. Zafar, A. Saadat, T.M. Khan, S. Iqbal, and I. Razzak, *LSSF-Net: Lightweight segmentation with self-awareness, spatial attention, and focal modulation.* Artificial Intelligence in Medicine, 2024: p. 103012.
6. Naveed, A., S.S. Naqvi, S. Iqbal, I. Razzak, H.A. Khan, and T.M. Khan, *RA-Net: Region-Aware Attention Network for Skin Lesion Segmentation.* Cognitive Computation, 2024: p. 1-18.
7. Khan, T.M., S.S. Naqvi, and E. Meijering, *ESDMR-Net: A lightweight network with expand-squeeze and dual multiscale residual connections for medical image segmentation.* Engineering Applications of Artificial Intelligence, 2024. **133**: p. 107995.
8. Mazher, M., I. Razzak, A. Qayyum, M. Tanveer, S. Beier, T. Khan, and S.A. Niederer, *Self-supervised spatial–temporal transformer fusion based federated framework for 4D cardiovascular image segmentation.* Information Fusion, 2024. **106**: p. 102256.
9. Naveed, A., S.S. Naqvi, T.M. Khan, S. Iqbal, M.Y. Wani, and H.A. Khan, *AD-Net: Attention-based dilated convolutional residual network with guided decoder for robust skin lesion segmentation.* Neural Computing and Applications, 2024: p. 1-23.
10. Iqbal, S., T.M. Khan, S.S. Naqvi, A. Naveed, M. Usman, H.A. Khan, and I. Razzak, *Ldmres-Net: a lightweight neural network for efficient medical image segmentation on iot and edge devices.* IEEE Journal of Biomedical and Health Informatics, 2023.
11. Naveed, A., S.S. Naqvi, T.M. Khan, and I. Razzak, *PCA: progressive class-wise attention for skin lesions diagnosis.* Engineering Applications of Artificial Intelligence, 2024. **127**: p. 107417.
12. Iqbal, S., T.M. Khan, S.S. Naqvi, and G. Holmes, *MLR-Net: A multi-layer residual convolutional neural network for leather defect segmentation.* Engineering Applications of Artificial Intelligence, 2023. **126**: p. 107007.
13. Khan, T.M., S.S. Naqvi, A. Robles-Kelly, and I. Razzak, *Retinal vessel segmentation via a Multi-resolution Contextual Network and adversarial learning.* Neural Networks, 2023. **165**: p. 310-320.
14. Iqbal, S., K. Naveed, S.S. Naqvi, A. Naveed, and T.M. Khan, *Robust retinal blood vessel segmentation using a patch-based statistical adaptive multi-scale line detector.* Digital Signal Processing, 2023. **139**: p. 104075.
15. Naqvi, S.S., Z.A. Langah, H.A. Khan, M.I. Khan, T. Bashir, M.I. Razzak, and T.M. Khan, *Glan: Gan assisted lightweight attention network for biomedical imaging based diagnostics.* Cognitive Computation, 2023. **15**(3): p. 932-942.
16. Qayyum, A., M. Mazher, T. Khan, and I. Razzak, *Semi-supervised 3D-InceptionNet for segmentation and survival prediction of head and neck primary cancers.* Engineering Applications of Artificial Intelligence, 2023. **117**: p. 105590.
17. Iqbal, S., T.M. Khan, K. Naveed, S.S. Naqvi, and S.J. Nawaz, *Recent trends and advances in fundus image analysis: A review.* Computers in Biology and Medicine, 2022. **151**: p. 106277.
18. Arsalan, M., T.M. Khan, S.S. Naqvi, M. Nawaz, and I. Razzak, *Prompt deep light-weight vessel segmentation network (PLVS-Net).* IEEE/ACM Transactions on Computational Biology and Bioinformatics, 2022. **20**(2): p. 1363-1371.
19. Yaqub, M., F. Jinchao, S. Ahmed, A. Mehmood, I.S. Chuhan, M.A. Manan, and M.S. Pathan, *DeepLabV3, IBCO-based ALCResNet: A fully automated classification, and grading system for brain tumor.* Alexandria Engineering Journal, 2023. **76**: p. 609-627.
20. Khan, A.Q., G. Sun, Y. Li, A. Bilal, and M.A. Manan, *Optimizing Fully Convolutional Encoder-Decoder Network for Segmentation of Diabetic Eye Disease.* Computers, Materials & Continua, 2023. **77**(2).
21. Raheem, A., Z. Yang, H. Yu, M. Yaqub, F. Sabah, S. Ahmed, . . . I.S. Chuhan, *IPC-CNN: A Robust Solution for Precise Brain Tumor Segmentation Using Improved Privacy-Preserving Collaborative Convolutional Neural Network.* KSII Transactions on Internet and Information Systems (TIIS), 2024. **18**(9): p. 2589-2604.
22. Manan, M.A., F. Jinchao, T.M. Khan, M. Yaqub, S. Ahmed, and I.S. Chuhan, *Semantic segmentation of

23. Manan, M.A., J. Feng, M. Yaqub, S. Ahmed, S.M.A. Imran, I.S. Chuhan, and H.A. Khan, *Multi-scale and multi-path cascaded convolutional network for semantic segmentation of colorectal polyps.* Alexandria Engineering Journal, 2024. **105**: p. 341-359.
24. Fang, Y., C. Chen, Y. Yuan, and K.-y. Tong. *Selective feature aggregation network with area-boundary constraints for polyp segmentation*. in *Medical Image Computing and Computer Assisted Intervention–MICCAI 2019: 22nd International Conference, Shenzhen, China, October 13–17, 2019, Proceedings, Part I 22*. 2019. Springer.
25. Zhou, Z., M.M.R. Siddiquee, N. Tajbakhsh, and J. Liang, *Unet++: Redesigning skip connections to exploit multiscale features in image segmentation.* IEEE transactions on medical imaging, 2019. **39**(6): p. 1856-1867.
26. Zhao, X., L. Zhang, and H. Lu. *Automatic polyp segmentation via multi-scale subtraction network*. in *Medical Image Computing and Computer Assisted Intervention–MICCAI 2021: 24th International Conference, Strasbourg, France, September 27–October 1, 2021, Proceedings, Part I 24*. 2021. Springer.
27. Fan, D.-P., G.-P. Ji, T. Zhou, G. Chen, H. Fu, J. Shen, and L. Shao. *Pranet: Parallel reverse attention network for polyp segmentation*. in *International conference on medical image computing and computer-assisted intervention*. 2020. Springer.
28. Dong, B., W. Wang, D.-P. Fan, J. Li, H. Fu, and L. Shao, *Polyp-pvt: Polyp segmentation with pyramid vision transformers.* arXiv preprint arXiv:2108.06932, 2021.
29. Bui, N.-T., D.-H. Hoang, Q.-T. Nguyen, M.-T. Tran, and N. Le, *MEGANet: Multi-Scale Edge-Guided Attention Network for Weak Boundary Polyp Segmentation.* arXiv preprint arXiv:2309.03329, 2023.
30. Liu, F., Z. Hua, J. Li, and L. Fan, *Dbmf: Dual branch multiscale feature fusion network for polyp segmentation.* Computers in Biology and Medicine, 2022. **151**: p. 106304.
31. Rahman, M.M. and R. Marculescu. *Medical image segmentation via cascaded attention decoding*. in *Proceedings of the IEEE/CVF Winter Conference on Applications of Computer Vision*. 2023.
32. Zhou, T., Y. Zhou, K. He, C. Gong, J. Yang, H. Fu, and D. Shen, *Cross-level Feature Aggregation Network for Polyp Segmentation.* Pattern Recognition, 2023. **140**: p. 109555.
33. Jain, S., R. Atale, A. Gupta, U. Mishra, A. Seal, A. Ojha, . . . O. Krejcar, *CoInNet: A Convolution-Involution Network with a Novel Statistical Attention for Automatic Polyp Segmentation.* IEEE Transactions on Medical Imaging, 2023.
34. Wei, J., Y. Hu, R. Zhang, Z. Li, S.K. Zhou, and S. Cui. *Shallow attention network for polyp segmentation*. in *Medical Image Computing and Computer Assisted Intervention–MICCAI 2021: 24th International Conference, Strasbourg, France, September 27–October 1, 2021, Proceedings, Part I 24*. 2021. Springer.
35. Tyagi, S., D.T. Kushnure, and S.N. Talbar, *An amalgamation of vision transformer with convolutional neural network for automatic lung tumor segmentation.* Computerized Medical Imaging and Graphics, 2023. **108**: p. 102258.
36. Theckedath, D. and R. Sedamkar, *Detecting affect states using VGG16, ResNet50 and SE-ResNet50 networks.* SN Computer Science, 2020. **1**: p. 1-7.
37. Bernal, J., F.J. Sánchez, G. Fernández-Esparrach, D. Gil, C. Rodríguez, and F. Vilariño, *WM-DOVA maps for accurate polyp highlighting in colonoscopy: Validation vs. saliency maps from physicians.* Computerized medical imaging and graphics, 2015. **43**: p. 99-111.
38. Jha, D., P.H. Smedsrud, M.A. Riegler, P. Halvorsen, T. de Lange, D. Johansen, and H.D. Johansen. *Kvasir-seg: A segmented polyp dataset*. in *MultiMedia Modeling: 26th International Conference, MMM 2020, Daejeon, South Korea, January 5–8, 2020, Proceedings, Part II 26*. 2020. Springer.
39. Ronneberger, O., P. Fischer, and T. Brox. *U-net: Convolutional networks for biomedical image segmentation*. in *Medical image computing and computer-assisted intervention–MICCAI 2015: 18th international conference, Munich, Germany, October 5-9, 2015, proceedings, part III 18*. 2015. Springer.
40. Badrinarayanan, V., A. Kendall, and R. Cipolla, *Segnet: A deep convolutional encoder-decoder architecture for image segmentation.* IEEE transactions on pattern analysis and machine intelligence, 2017. **39**(12): p. 2481-2495.
41. Liu, W., A. Yue, W. Shi, J. Ji, and R. Deng. *An automatic extraction architecture of urban green space based on DeepLabv3plus semantic segmentation model*. in *2019 IEEE 4th International Conference on Image, Vision and Computing (ICIVC)*. 2019. IEEE.